\documentclass[11pt]{article}
\usepackage{epsf} 
\usepackage{times}
\usepackage{bm} 
\usepackage{graphics}
\usepackage[round]{natbib}

\begin{document} 

\def\tr{{\mathrm T}} \def\rd{{\mathrm d}} \def\e{{\mathrm e}}
\def\i{{\mathrm i}} \def\ssum{{\mbox{\small{$\Sigma$}}}}
\newcommand{\mb}[1]{\ensuremath{\bm{#1}}} 

\author{Magnus Rattray \\ \ \\ 
Computer Science Department, 
University of Manchester, \\
Manchester M13 9PL, United Kingdom.\\ \ \\} 


\date{To appear in Neural Computation}
 
\title{Stochastic Trapping in a Solvable Model of On-line Independent
Component Analysis} 

\maketitle

\section*{Abstract}

Previous analytical studies of on-line Independent Component Analysis
(ICA) learning rules have focussed on asymptotic stability and
efficiency. In practice the transient stages of learning will often be
more significant in determining the success of an algorithm. This is
demonstrated here with an analysis of a Hebbian ICA algorithm which
can find a small number of non-Gaussian components given data composed
of a linear mixture of independent source signals. An idealised data
model is considered in which the sources comprise a number of
non-Gaussian and Gaussian sources and a solution to the dynamics is
obtained in the limit where the number of Gaussian sources is
infinite. Previous stability results are confirmed by expanding around
optimal fixed points, where a closed form solution to the learning
dynamics is obtained. However, stochastic effects are shown to
stabilise otherwise unstable sub-optimal fixed points.  Conditions
required to destabilise one such fixed point are obtained for the case
of a single non-Gaussian component, indicating that the initial
learning rate $\eta$ required to successfully escape is very low
($\eta = O(N^{-2})$ where $N$ is the data dimension) resulting in very
slow learning typically requiring $O(N^3)$ iterations. Simulations
confirm that this picture holds for a finite system.

\newpage

\section{Introduction} 

Independent component analysis (ICA) is a statistical modelling
technique which has attracted a significant amount of
research interest in recent years \citep[for a review,
see][]{hyva99}. In ICA the goal is to find a representation of
data in terms of a combination of statistically independent
variables. This technique has a number of
useful applications, most notably blind source separation,
feature extraction and blind deconvolution. A large number of neural
learning algorithms have been applied to this problem, as detailed in the
aforementioned review. 

Theoretical studies of on-line ICA algorithms have
mainly focussed on asymptotic stability and efficiency, using the
established results of stochastic approximation theory. However,
in practice the transient stages of learning will often be more significant in
determining the success of an algorithm. In this paper a Hebbian ICA 
algorithm is analysed and a solution to
the learning dynamics is obtained in the limit of large data dimension. The analysis
highlights the critical
importance of the transient dynamics and in particular an extremely
low initial learning rate is found to be essential in order to avoid
trapping in a sub-optimal fixed point close to the initial conditions
of the learning dynamics. 

This work focuses on the bigradient learning algorithm
introduced by \citet{wang96} and studied in the context of ICA by
\citet{hyva98} where it was shown to have nice
stability conditions. This algorithm can be used to extract a small
number of independent components from data of high dimension and is
closely related to projection pursuit algorithms which detect
``interesting'' projections in high-dimensional data. The algorithm can be
defined in on-line mode or can form the basis of a
fixed-point batch algorithm which has been found to improve computational
efficiency~\citep{hyva97}. In this work the dynamics of the
on-line algorithm is studied. This may be the preferred mode when the
model parameters are non-stationary or when the data set is very
large. Although the analysis is restricted to a stationary data model, the results are
relevant to the non-stationary case in which learning strategies are
often designed to increase the learning rate when far from the
optimum~\citep{muller98}. The results obtained here suggest that this strategy can
lead to very poor performance.

In order to gain insight into the learning dynamics an
idealised model is considered in which data is composed of a small number of
non-Gaussian source signals linearly mixed in with a large number of
Gaussian signals. A solution to the dynamics is obtained
in the limiting case where the number of Gaussian signals is
infinite. In this limit one can use techniques from statistical
mechanics similar to those which have previously been applied to other on-line learning
algorithms, including other unsupervised Hebbian learning algorithms such as Sanger's
PCA algorithm~\citep{biehl98}. For the asymptotic dynamics close to an optimal solution the
stability conditions due to~\citet{hyva98}
are confirmed and the eigensystem is obtained which determines the
asymptotic dynamics and
optimal learning rate decay. However, the dynamical equations also
have sub-optimal fixed points which are stable for any
$O(N^{-1})$ learning rate where $N$ is the data dimension. Conditions required to
destabilise one such fixed point are obtained in the case of a single 
non-Gaussian source, indicating that learning must be very
slow initially in order to learn successfully. The analysis requires a careful treatment 
of fluctuations which prove to be important even in the limit of large input dimension. Finally, 
simulation results are presented which
indicate that this phenomenon persists also in finite sized systems.

\section{Data model} 

\begin{figure}[h]
	\setlength{\unitlength}{0.6cm}
	\begin{center}
	\begin{picture}(10,10)
	\epsfysize = 9cm
	\put(1,0){\epsfbox[50 0 550 600]{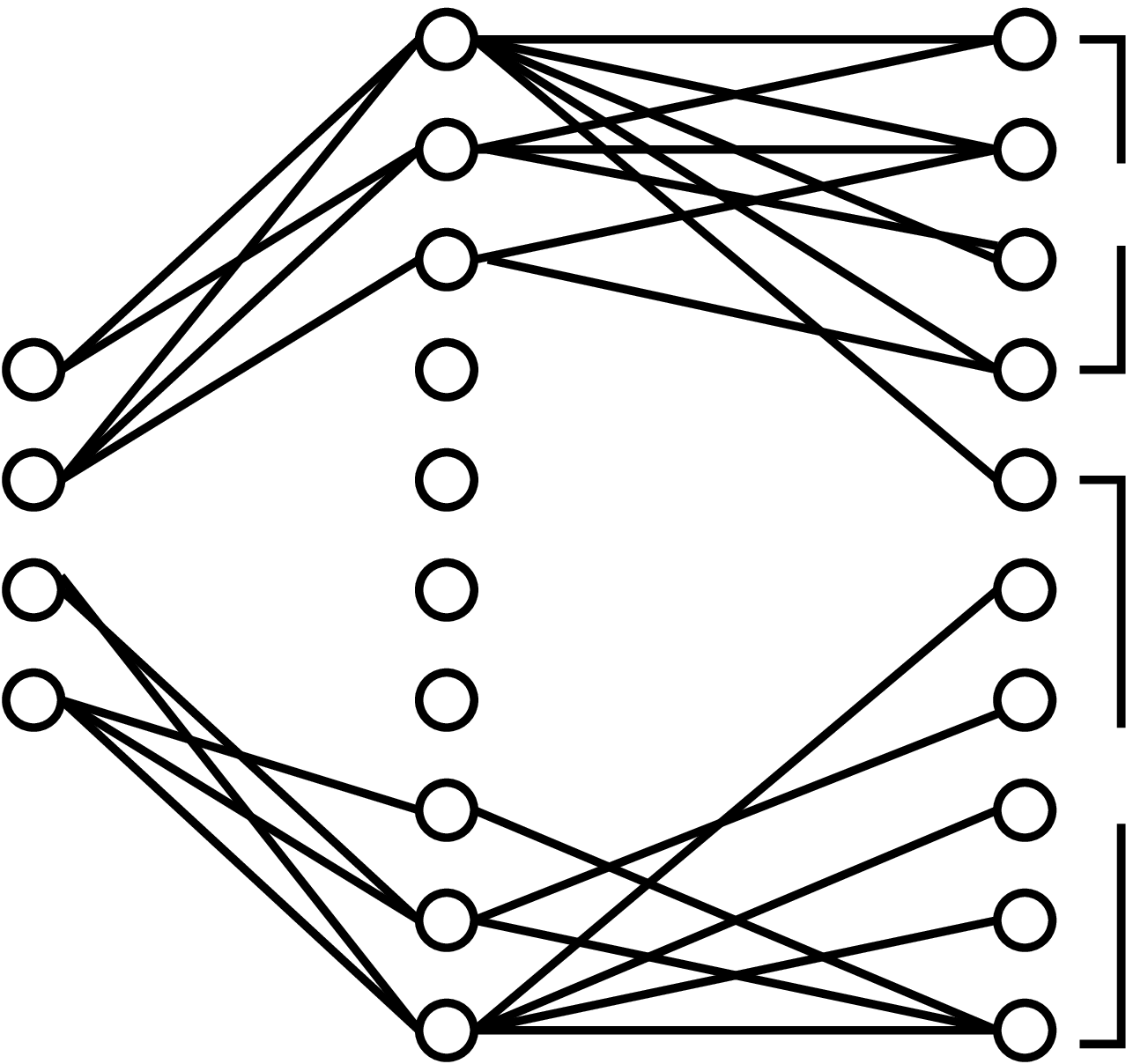}}
	\put(-2.6,1.4){\large{\mbox{$\mathsf{\bm y=\bm W^\tr \bm x}$}}}
	\put(1.3,4){\Large{\mbox{$\mathsf{\bm W}$}}}
	\put(1.1,-1.0){\large{\mbox{$\mathsf{\bm x=\bm A_s \bm
	s + \bm A_n \bm n}$}}}
	\put(4.3,4){\Large{\mbox{$\mathsf{[\bm A_s \; \bm A_n]}$}}}
	\put(8.65,6.85){\large{\mbox{$\mathsf{\bm s}$}}}
	\put(8.65,2.2){\large{\mbox{$\mathsf{\bm n}$}}}
	\end{picture}
	\end{center}
        \caption{The linear mixing model generating the data is shown above.
There are $M$ non-Gaussian independent sources $\bm{s}$ while the
remaining $N-M$ sources $\bm{n}$ are uncorrelated Gaussian variables.
There are $N$ outputs $\bm{x}$ formed by multiplying the sources by the
square non-singular mixing matrix ${\bm A}\equiv [\bm A_s \, \bm
A_n]$. The outputs are linearly
projected onto the $K$-dimensional vector $\bm{y}=\bm{W}^\tr\bm{x}$. 
It is assumed that $M\ll N$ and $K\ll N$.}
        \label{fig_network} 
\end{figure}

The linear data model is shown in figure~\ref{fig_network}.
In order to apply the Hebbian ICA algorithm one should first sphere the
data, ie. linearly transform the data so that it has an identity
covariance matrix. This can be achieved by standard transformations in a batch
setting but for on-line learning an adaptive sphering algorithm, such 
as the one introduced by~\citet{cardoso96}, could be
used. To simplify matters it is assumed here that the data has
already been sphered. Without loss of generality it can also be assumed 
that the sources each have unit variance. The sources are decomposed
into $M$ non-Gaussian
and $N-M$ Gaussian components respectively,
\begin{equation}
	p(\bm s)=\prod_{i=1}^M p_i(s_i) \ , \qquad p(\bm n) =
\prod_{i=M+1}^N \frac{\e^{\frac{-n_i^2}{2}}}{\sqrt{2\pi}} \ .
\end{equation}
To conform with the above model assumptions the mixing matrix $\bm A$ must be
unitary. The mixing matrix is decomposed into two rectangular matrices
$\bm A_s$ and $\bm A_n$ associated with the non-Gaussian and Gaussian
components respectively, 
\begin{equation}
 \bm x = \bm A \left[\begin{array}{c} \bm s
\\ \bm n \end{array}\right] = \left[\bm A_s \; \bm
A_n\right]\left[\begin{array}{c} \bm s \\ \bm n \end{array}\right] = \bm
A_s \bm s + \bm A_n \bm n \ . 
\label{eqn_x}
\end{equation}
The unitary nature of $\bm A$ results in the following constraints, 
\begin{eqnarray} \left[\bm A_s \; \bm
A_n\right]\left[\begin{array}{c} \bm A_s^\tr \\ \bm A_n^\tr
\end{array}\right] & = & \bm A_s \bm A_s^\tr + \bm A_n \bm A_n^\tr = \bm
I \ , \label{eqn_Ac1} \\ \left[\begin{array}{c} \bm A_s^\tr \\ \bm
A_n^\tr \end{array}\right]\left[\bm A_s \; \bm A_n\right] & = &
\left[\begin{array}{c c} \bm A_s^\tr \bm A_s & \bm A_s^\tr \bm A_n \\
\bm A_n^\tr \bm A_s & \bm A_n^\tr \bm A_n \end{array}\right] =
\left[\begin{array}{c c} \bm I & \bm 0 \\ \bm 0 & \bm I
\end{array}\right] \label{eqn_Ac2} \ . 
\end{eqnarray} 

\section{Algorithm}

The following on-line Hebbian learning rule was introduced by~\citet{wang96} 
and analysed in the context of ICA by~\citet{hyva98},
\begin{equation} 
\bm W^{t+1} - \bm W^t = \eta\,\bm\sigma\bm x^t
\bm\phi(\bm y^t)^\tr + \alpha \bm W^t(\bm I - (\bm W^t)^\tr\bm W^t) \ ,
\label{eqn_dW}
\end{equation} 
where $\bm\phi(\bm y^t)_i=\phi(y^t_i)$ is an odd
non-linear function which is applied to every component of the
$K$-dimensional vector $\bm y\equiv\bm W^\tr\bm
x$. The first term on the right
is derived by maximising the non-Gaussianity of the projections in
each component of the vector $\bm y$. The second term ensures that the columns of $\bm W$
are close to orthogonal so that the projections are uncorrelated and of
unit variance. The learning rate
$\eta$ is a positive scalar parameter which must be chosen 
with care and may depend on time. The parameter $\alpha$ is less
critical and setting $\alpha=0.5$ seems to
provide reasonable performance in general. The diagonal matrix $\bm \sigma$ has elements
$\sigma_{ii}\in\{-1,1\}$ which ensure the stability of the desired fixed
point. The elements of this matrix can either be chosen adaptively or
can be chosen according to {\it {\'a} priori} knowledge about the source
statistics. Stability of the optimal fixed point is ensured by the
condition~\citep{hyva98}, 
\begin{equation} 
\sigma_{ii} = \mbox{Sign}(\langle s_i\phi(s_i)-\phi'(s_i)\rangle) \ , 
\label{eqn_stability}
\end{equation}
assuming we order indices such that $y_i\rightarrow \pm s_i$ for $i\leq\min(K,M)$
asymptotically. The angled brackets denote an average over the source
distribution. 

A remarkable feature of the above algorithm is that the same
non-linearity can be used for source signals with very different
characteristics. For example, both sub-Gaussian and super-Gaussian
signals can be separated using either $\phi(y)=y^3$ or $\phi(y)=\tanh(y)$, two
common choices of non-linearity. 

\section{Dynamics for large input dimension} 

Define the following two matrices, 
\begin{equation} 
\bm R \equiv \bm W^\tr \bm A_s \ , \qquad \bm Q \equiv \bm W^\tr \bm W
\ . 
\label{eqn_RQ}
\end{equation}
Using the constraint in equation~(\ref{eqn_Ac1}) one can show that, 
\begin{eqnarray} 
\bm y & = & \bm W^\tr(\bm A_s \bm s + \bm A_n \bm n) \nonumber \\ & =
& \bm R \bm s + \bm
z \quad \mbox{ where } \quad \bm z \sim {\cal{N}}(\bm 0,\bm Q -
\bm{RR}^\tr) \ . 
\label{eqn_y}
\end{eqnarray} 
Knowledge of the matrices $\bm R$ and $\bm Q$ is
therefore sufficient to describe the relationships between the
projections $\bm y$ and the sources $\bm s$ in full. Although the
dimension of the data is $N$, the dimension of these matrices is 
$K\times M$ and $K\times K$ respectively. The
system can therefore be described by a small number of macroscopic quantities in the limit of large
$N$ as long as $K$ and $M$ remain manageable. 

In appendix~\ref{app_dynamics} it is shown that in the limit $N \rightarrow
\infty$, $\bm Q\rightarrow\bm I$ while $\bm R$ evolves deterministically
according to the following first order differential equation,
\begin{equation} 
\frac{\rd \bm R}{\rd \tau} =
\mu\bm\sigma\left(\langle \bm\phi(\bm y)\bm s^\tr\rangle -
\mbox{$\frac{1}{2}$}\langle \bm\phi(\bm y)\bm y^\tr + \bm y\bm\phi(\bm
y)^\tr\rangle\bm R\right) - \mbox{$\frac{1}{2}$}\mu^2\langle\bm\phi(\bm
y)\bm\phi(\bm y)^\tr\rangle \bm R \label{eqn_dRdt} 
\end{equation} 
with rescaled variables $\tau\equiv t/N$ and $\mu\equiv N\eta$. This 
deterministic equation is only valid for $\bm R= O(1)$ and a different 
scaling is considered in section~\ref{sec_escape}, in which case 
fluctuations have to be considered even in the limit. The brackets denote
expectations with respect to the source distribution and $\bm z\sim
{\cal{N}}(\bm 0,\bm I - \bm{RR}^\tr)$. The
bracketed terms therefore only depend on $\bm R$ and statistics of the
source distributions, so that the above equation forms a closed system.

\subsection{Optimal asymptotics}
\label{sec_asympt}

The desired solution is one where as many as possible of the $K$ 
projections mirror one of the $M$ sources. If $K<M$
then not all the sources can be learned and which ones are learned 
depends on details of the initial conditions. If $K\geq M$ then which
projections mirror each source also depends on the
initial conditions. For $K>M$ there will be projections which do
not mirror any sources; these will be statistically independent of the
sources and have a Gaussian distribution with identity
covariance matrix.

Consider the case where $y_i\rightarrow s_i$ for $i=1\ldots \min(K,M)$
asymptotically (all other solutions can be obtained by a
trivial permutation of indices and/or changes in sign). The optimal solution
is then given by
$R^*_{ij} =\delta_{ij}$ which is a fixed point of
equation~(\ref{eqn_dRdt}) as $\mu\rightarrow 0$. Asymptotically the learning rate 
should be annealed in order to approach this fixed point and the usual inverse law 
annealing can be shown to be optimal subject to a good choice of
prefactor,
\begin{equation}
	\mu \sim \frac{\mu_0}{\tau} \ \ \mbox{ as } \ \ \tau\rightarrow \infty \  
\quad \left(\mbox{ or equivalently } \eta \sim \frac{\mu_0}{t} \ \ \mbox{ as } \ \ t\rightarrow \infty\right) \ .
\end{equation}
The asymptotic 
solution to equation~(\ref{eqn_dRdt}) with the above annealing
schedule was given by~\cite{leen98}. Let $u_{ij}=R_{ij}-R^*_{ij}$ be
the deviation from the fixed point. Expanding equation~(\ref{eqn_dRdt})
around the fixed point one obtains,
\begin{equation}
	u_{ij}(\tau) \sim
         \sum_{k,n=1}^K\sum_{l,m=1}^M
         V_{ijkl}\left\{ -\mbox{$\frac{1}{2}$}X_{kl}\langle\phi^2(s_n)\rangle\delta_{nm}+\left(\frac{\tau_0}{\tau}\right)^{-\mu_0\lambda_{kl}}\!\!\!\!\!\!u_{nm}(\tau_0)\right\}V^{-1}_{klnm} \ , 
\label{eqn_asymp}
\end{equation}
where,
\begin{equation}
	X_{ij} = \left(\frac{\mu_0^2}{-\mu_0\lambda_{ij}-1}\right)\left[\frac{1}{\tau}-\frac{1}{\tau_0}\left(\frac{\tau_0}{\tau}\right)^{-\mu_0\lambda_{ij}}\right]  \ . 
\label{eqn_X} 
\end{equation}
Here, $\tau_0$ is the time at which annealing begins and $\lambda_{ij}$ and $V_{ijkl}$ are the
eigenvalues and eigenvectors of the
Jacobian of the learning equation to first order in $\mu$. These are
written as matrices and tensors respectively rather than
scalars and vectors because the system's variables are in a matrix. One
can think of pairs of indices $(i,j)$ and $(k,l)$ as each representing a single
index in a vectorised system. The explicit solution to the 
eigensystem is given in appendix~\ref{app_eigensystem}. From the
eigenvalues, defined in equation~(\ref{eqn_evals}), it is clear that
the fixed point is stable if and only if the condition in
equation~(\ref{eqn_stability}) is met. 

There is a critical learning
rate, $\mu_0^{\mbox{\tiny crit}}=-1/\lambda_{\mbox{\tiny max}}$, where
$\lambda_{\mbox{\tiny max}}$ is the largest eigenvalue (smallest
in magnitude, since the eigenvalues are negative), such that if
$\mu_0<\mu_0^{\mbox{\tiny crit}}$ then the approach to the fixed point will be
slower than the optimal $1/\tau$ decay. From
the eigenvalues given in~(\ref{eqn_evals}) we find that $\lambda_{\mbox{\tiny
max}}=-\xi_{\mbox{\tiny min}}$ where $\xi_i=\sigma_{ii}\langle
s_i\phi(s_i) - \phi'(s_i)\rangle$, so that $\mu_0^{\mbox{\tiny
crit}}=1/\xi_{\mbox{\tiny min}}$. As long as
$\mu_0>\mu_0^{\mbox{\tiny crit}}$ then the terms involving $\tau_0$ in
equations~(\ref{eqn_asymp}) and (\ref{eqn_X}) will be negligible and the asymptotic decay will
be independent of the initial conditions and transient dynamics.

Assuming $\mu_0>\mu_0^{\mbox{\tiny crit}}$ and substituting in the explicit
expressions for the eigensystem we find the following simple form for the asymptotic
dynamics to leading order in $1/\tau$,
\begin{equation}
  u_{ij}(\tau) \sim
  -\frac{\delta_{ij}}{\tau}\left(\frac{\mu_0^2\langle \phi^2(s_i)\rangle}{4\mu_0\xi_{i}-2}\right) \ .
\label{eqn_asympt2}
\end{equation}

\subsection{Escape from the initial transient} 
\label{sec_escape}

Unfortunately, the optimal fixed points described in the previous
section are not the only stable fixed points of
equation~(\ref{eqn_dRdt}). In some cases the algorithm will converge
to a sub-optimal solution in which one or more potentially detected
signals remain unlearned and the corresponding entry in $\bm R$ decays
to zero. The stability of these points is due to the $O(\mu^2)$ term in
equation~(\ref{eqn_dRdt}) which becomes less significant as the learning
rate is reduced, in which case the corresponding negative eigenvalue
of the Jacobian eventually vanishes. Higher order terms then lead
to instability and escape from this sub-optimal fixed point. One can
therefore avoid trapping by selecting a sufficiently low learning rate
during the initial transient. 

Consider the simplest case where $K=M=1$ in which case the matrix 
$\bm R$ reduces to a scalar: $R=R_{11}$ 
and $\sigma=\sigma_{11}$. Expanding
equation~(\ref{eqn_dRdt}) around $R=0$,
\begin{equation}
	\frac{\rd R}{\rd \tau} = -\mbox{$\frac{1}{2}$}
\langle\phi(z)^2\rangle\, \mu^2R  + \mbox{$\frac{1}{6}$}\kappa_4 \langle
\phi'''(z)\rangle\, \sigma\mu R^3 + O(R^5)\ .
\label{eqn_smallR}
\end{equation}
Here, $\kappa_4$ is the fourth cumulant of the source
distribution and the brackets denote averages over $z\sim {\cal
N}(0,1)$. Although $R=0$ is a stable 
fixed point, the range of attraction is reduced as $\mu\rightarrow 0$
until eventually instability occurs. The condition under which one will
successfully escape the fixed point is found by setting $\rd |R|/\rd t>0$,
\begin{equation}
	\sigma = \mbox{Sign}(\kappa_4\langle\phi'''(z)\rangle) \ , \quad
\mu<\frac{R^2 |\langle\phi'''(z) \rangle
\kappa_4 |}{3\langle\phi^2(z)\rangle} \ .
\label{eqn_escape}
\end{equation}

Notice that the condition on $\sigma$ for escaping the
initial transient is not generally the same as the condition in
equation~(\ref{eqn_stability}) which ensures stability of the asymptotic fixed point. For
$\phi(y)=y^3$ the conditions are exactly equivalent. However, in other cases
the conditions may conflict and an adaptive choice of $\sigma$ based
on equation~(\ref{eqn_stability}), as suggested
by~\cite{hyva98}, may give poor results. With $\phi(y) = \tanh(y)$
the conditions appear to be equivalent in many cases. This condition
is equally applicable to gradient based batch algorithms since it is
due to the $O(\mu)$ term above, which is not related to fluctuations.

If the entries in $\bm A$ and $\bm W$ are initially of similar order then one would
expect $R= O(N^{-\frac{1}{2}})$. This is the typical case if we
consider a random and uncorrelated choice for $\bm A$ and the initial
entries in $\bm W$. Larger initial values of $R$ could only be obtained with some prior
knowledge of the mixing matrix which we will not assume. With $R=
O(N^{-\frac{1}{2}})$ the initial
value of $\mu$ required to escape is $O(N^{-1})$, indicating a very slow initial
phase in the dynamics (recall that the unscaled learning rate
$\eta\equiv\mu/N$). Larger learning rates will result in trapping in
the sub-optimal fixed point. However, the above
result is not strictly valid unless $R= O(1)$, since this was an
assumption used in the derivation of
equation~(\ref{eqn_dRdt}). When $R= O(N^{-\frac{1}{2}})$ then one can
no longer assume that fluctuations are negligible as $N\rightarrow
\infty$.  Define the $O(1)$ quantities $r\equiv R\sqrt{N}$ and $\nu\equiv
\eta N^2$. The mean and variance of the change in $r$ at each learning
step can be calculated (to leading order in $N^{-1}$),
\begin{eqnarray}
	\mbox{E}[\Delta r] & \simeq & \left( -
	\mbox{$\frac{1}{2}$}\langle\phi^2(z)\rangle\nu^2 r + \mbox{$\frac{1}{6}$}\kappa_4\langle \phi'''(z)
	\rangle\sigma\nu \, r^3\right)N^{-3} \ , \label{eqn_dr}\\
	\mbox{Var}[\Delta r] & \simeq & \langle
	\phi^2(z)\rangle \nu^2 N^{-3} \ .
\label{eqn_diff}
\end{eqnarray}
This expression is derived by a similar adiabatic elimination of $Q_{11}$ as
carried out for the deterministic case in
appendix~\ref{app_dynamics}. This requires that $\alpha$ and $\eta$
are the same order before taking the limit $N\rightarrow\infty$,
followed by the limit $\alpha\rightarrow\infty$ and corresponds to
the usual form of ``slaving'' in Haken's terminology in which the
eliminated variable only contributes to the change in mean. Other
scalings may result in slightly different
expressions~\citep[see, for example,][section 6.6]{gardiner} although it is expected that the
main conclusions described below will not be affected.

\begin{figure}[h]
	\setlength{\unitlength}{0.5cm}
	\begin{center}
	\begin{picture}(15,8)
	\epsfysize = 6cm
	\put(0,-3){\epsfbox[50 100 550 600]{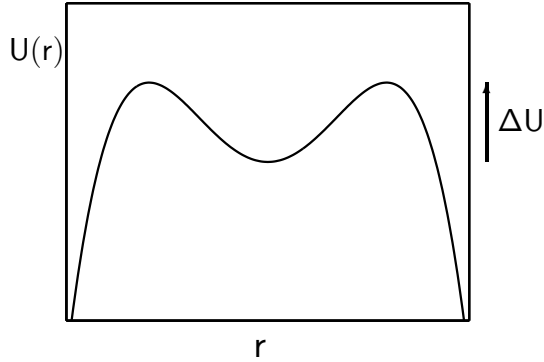}}
	\put(-0.5,7){\large{\mbox{$\mathsf{U(r)}$}}}
	\put(6,-0.9){\Large{\mbox{$\mathsf{r}$}}}
	\put(12.2,4.3){\linethickness{1pt}{\vector(0,1){2.1}}}
	\put(12.5,5.1){\large{\mbox{$\mathsf{\Delta U}$}}}
	\end{picture}
	\end{center}
        \caption{For $r=R\sqrt{N}= O(1)$ the dynamics can be represented as
a diffusion in a symmetric quartic potential $U(r)$. The escape time from an
unstable fixed point at $r=0$ is mainly determined by the potential barrier
$\Delta U$.}
        \label{fig_potential} 
\end{figure}

The equation for the mean is similar to
equation~(\ref{eqn_smallR}). However, the variance is the same order
as the mean in the limit $N\rightarrow\infty$ and fluctuations cannot 
be ignored in this case. The system is better described by a
Fokker-Planck equation~\citep[see, for example,][]{gardiner} with a
characteristic time-scale of $O(N^3)$. The system is locally equivalent
to a diffusion in the following quartic potential,
\begin{equation}
	U(r) = \mbox{$\frac{1}{4}$}\langle \phi^2(z) \rangle \nu^2 r^2 -
	\mbox{$\frac{1}{24}$}|\kappa_4 \langle
	\phi'''(z)\rangle| \nu \, r^4 \ ,
\end{equation}
with a diffusion coefficient $D=\langle\phi^2(z)\rangle \nu^2$ which is
independent of $r$. The shape of this
potential is shown in figure~\ref{fig_potential}. A potential
barrier $\Delta U$ must be overcome to escape an unstable state
close to $R=0$ (assuming that the condition on $\sigma$ in
equation~(\ref{eqn_escape}) is satisfied).

For large $\nu$ this system corresponds to an Ornstein-Uhlenbeck
process with a Gaussian
stationary distribution of fixed unit variance. Thus, if one chooses
too large $\nu$ initially the dynamics will become localised close to $R=0$. As $\nu$ is
reduced the potential barrier confining the dynamics is reduced. The
time-scale for escape for large $\nu$ is mainly determined by the effective size
of the barrier~\citep{gardiner},
\begin{equation}
	T_{\mbox{\tiny escape}} \: \propto \: \exp\left(\frac{\Delta
	U}{D}\right) \: = \: \exp\left(\frac{3\langle\phi^2(z)\rangle\nu}{8|\kappa_4\langle\phi'''(z)\rangle
	|}\right)
	\ .
\end{equation}
As the learning rate is reduced so the time-scale for
escape is also reduced. However, the choice of optimal learning rate is
non-trivial and cannot be determined by considering only the
leading order terms in $R$ as above, because although small $\nu$ will
result in a quicker escape from the unstable fixed point near $R=0$
this will then lead to a very slow learning transient after escape. 

From the above discussion one can draw two important
conclusions. Firstly, the initial learning
rate should be $O(N^{-2})$ or less initially in order to avoid
trapping close to the initial conditions. Secondly, the time-scale required
to escape the initial transient is $O(N^3)$, resulting in an extremely
slow initial stage of learning. 

\subsection{Other sub-optimal fixed points} 
\label{sec_other}

In studies of other on-line learning algorithms, such as Sanger's rule
and back-propagation, a class of sub-optimal fixed points have been
discovered which are due to symmetries inherent in the learning
machine's structure~\citep{saad95a,saad95b,biehl95,biehl98}. These
symmetric fixed points are unstable for small learning rates, but the
eigenvalues determining escape are typically of very small magnitude
so that trapping can occur if the initial conditions are sufficiently
symmetric. In practice this will typically occur only for very large
input dimensions ($N>10^6$) and will result in learning timescales of
$O(N^2)$ for $O(N^{-1})$ learning rates. Equation~(\ref{eqn_dRdt})
does exhibit fixed points of this type for particular initial
conditions. Consider the case $K=M=2$ as an example. If initially
$R_{11}\simeq R_{21}$ and $R_{12}\simeq R_{22}$ then the dynamics will
preserve this symmetry until instabilities due to slight initial
differences lead to escape from an unstable fixed point. This
symmetry breaking is necessary for good performance since each
projection must specialise to a particular source signal.

As mentioned above, sufficiently small differences in the initial
value of the entries in $\bm R$ will typically only occur for very
large $N$, much larger than the typical values currently used in
ICA. A very small learning rate is then required to avoid trapping in
a fixed point near the initial conditions, as discussed in the
previous section. This initial trapping is far more serious
than the symmetric fixed point since it requires a learning rate of
$O(N^{-2})$ in order to escape, resulting in a far greater loss of
efficiency. In practice, symmetric fixed points do not appear to
be a serious problem and we have not observed any such fixed points in simulations
of finite systems. This may be due to the highly stochastic nature of
the initial dynamics, in which fluctuations are large compared to the average
dynamical trajectory. This is in contrast to the picture for
back-propagation, for example, where fluctuations result in
relatively small corrections to the average trajectory~\citep{barber96}. The strong fluctuations
observed here may help break any symmetries which might otherwise lead to
trapping in a symmetric fixed point, although a full understanding of
this effect requires careful analysis of the multivariate diffusion equation
describing the dynamics near the initial conditions.

\section{Simulation results}

The theoretical results in the previous section are for the limiting
case where $N\rightarrow\infty$. In practice we should verify that
the results are relevant in the case of large but finite $N$. In this
section simulation evidence is presented which demonstrates that the 
trapping predicted by the theory occurs in finite systems. 

\begin{figure}[h]
\setlength{\unitlength}{1.0cm}
	\begin{center}
	\begin{picture}(15,15)
	\epsfysize = 13cm
	\put(0,0){\epsfbox[50 100 550 600]{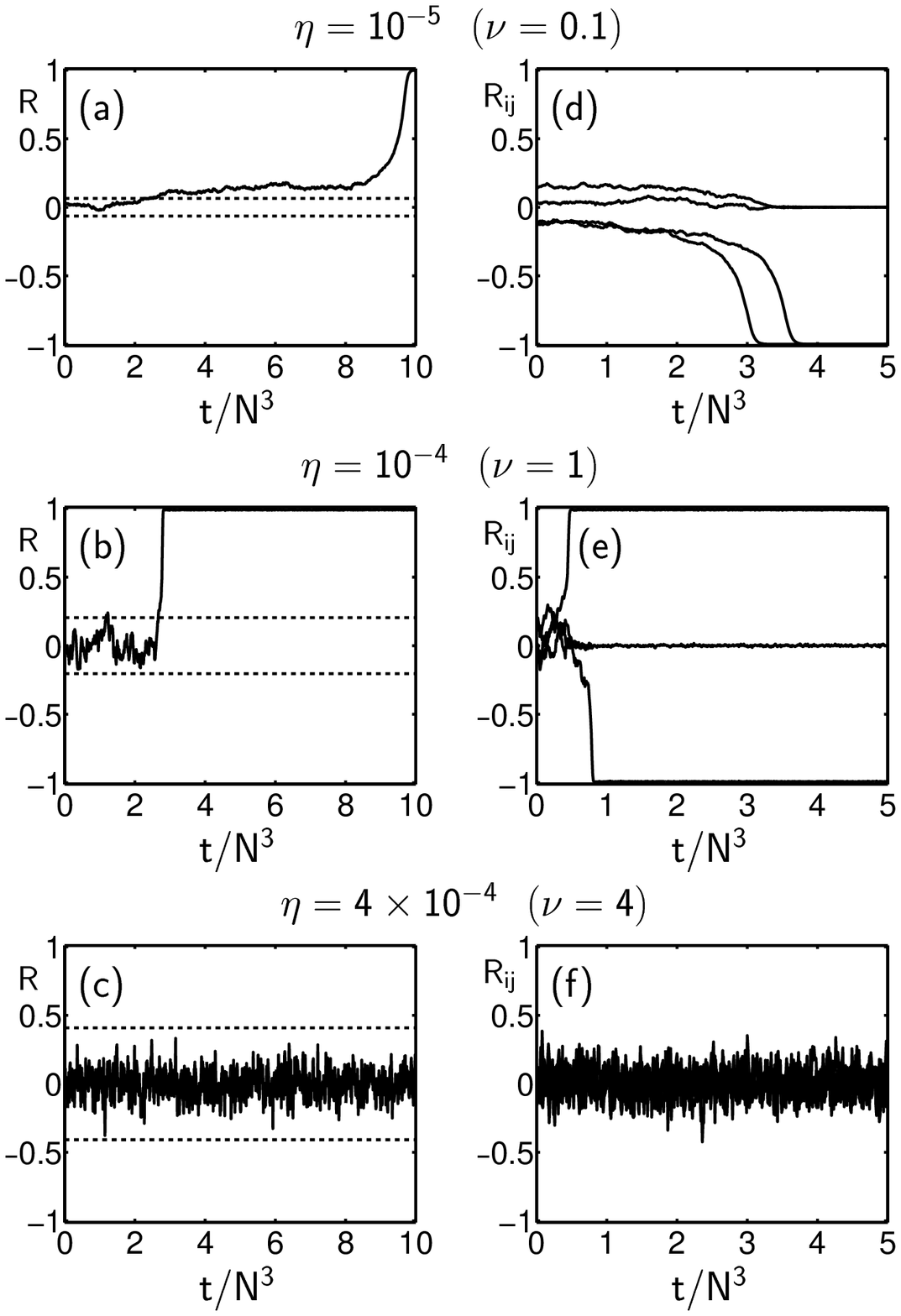}}
	\end{picture}
	\end{center}
        \caption{100-dimensional data ($N=100$) is produced from a mixture
containing a small number of uniformly distributed sources. Figures on the left (a--c) are for
a single non-Gaussian source and a single projection ($M=K=1$)
while figures on the right (d--f) are for two non-Gaussian
sources and two projections ($M=K=2$). Each column shows examples 
of learning with the same initial conditions and data but with different 
learning rates. From top to bottom: $\eta=10^{-5}\,(\nu=0.1)$, 
$\eta=10^{-4}\,(\nu=1)$ and $\eta=4\times 10^{-4}\,(\nu=4)$.}
       \label{fig_MK1} 
\end{figure}

Figures~\ref{fig_MK1}(a)--(c) show results produced by an algorithm learning a
single projection from 100-dimensional data with a single non-Gaussian
(uniformly distributed) component ($N=100,M=K=1$). The matrices $\bm
A$ and $\bm W$ are randomly initialised with orthogonal, normalised
columns. Similar results are obtained for
other random initialisations. A cubic non-linearity is used and
$\sigma$ is set to $-1$, although the adaptive scheme for setting
$\sigma$ suggested by~\citet{hyva98} gives similar results. In each
example, dashed lines show the maxima of the potential in
figure~\ref{fig_potential}. Figure~\ref{fig_MK1}(a) shows the learning
dynamics from a single run with $\eta=10^{-5}\,(\nu=0.1)$. The
dynamics follows a relatively smooth trajectory in this case and much
of the learning is determined by the cubic term in
equation~(\ref{eqn_dr}). With this choice of learning rate there is a
strong dependence on the initial conditions, with larger initial
magnitude of $R$ often resulting in significantly faster
learning. However, recall that a high value for $R$ cannot be chosen
without prior knowledge of the mixing matrix. Figure~\ref{fig_MK1}(b) shows the learning dynamics with a larger learning rate
$\eta=10^{-4}\,(\nu=1)$ for exactly the same initial conditions and
sequence of data. In this case the learning trajectory is more
obviously stochastic and is initially confined within the unstable
sub-optimal state with $R\simeq 0$. Eventually the system leaves this
unstable state and quickly approaches $R\simeq 1$. In this case the
dynamics is not particularly sensitive to the initial magnitude of $R$
although the escape time can vary significantly due to the inherent
randomness of the learning process. In figure~\ref{fig_MK1}(c) the
learning dynamics is shown for a larger learning rate $\eta=4\times
10^{-4}\,(\nu=4)$. In this case the system remains trapped in the
sub-optimal state for the entire simulation time.

The analysis in section~\ref{sec_escape} is only strictly valid for the
case of a single non-Gaussian source and a single projection. However, similar trapping occurs 
in general as demonstrated in figures~\ref{fig_MK1}(d)--(f). The components of $\bm R$
are plotted for an algorithm learning two projections 
from 100-dimensional data with two
non-Gaussian (uniformly distributed) components ($N=100,M=K=2$). The 
different learning regimes identified in the single component case are mirrored 
closely in the case of this two component model. 

\section{Conclusion} 

An on-line Hebbian ICA algorithm was studied for the case in which data comprises a 
linear mixture of Gaussian and non-Gaussian sources and a solution to the dynamics was 
obtained for the idealised scenario in which the number of non-Gaussian sources is
finite while the number of Gaussian sources is infinite. The analysis confirmed the stability conditions
found by~\citet{hyva98} and the eigensystem characterising the 
asymptotic regime was determined. However, it was also 
shown that there exist
sub-optimal fixed points of the learning dynamics which are stabilised by 
stochastic effects under certain conditions. The simplest case of a single 
non-Gaussian component was studied in detail. The analysis revealed that typically a
very low learning rate ($\eta= O(N^{-2})$ where $N$ is the data
dimension) is required to escape this sub-optimal fixed
point, resulting in a long learning time of $O(N^{3})$ iterations. Simulations of a 
finite system support these theoretical conclusions.

The sub-optimal fixed point studied here has some interesting features. In the limit
$\eta\rightarrow 0 $ the dynamics becomes
deterministic and fluctuations due to the stochastic nature of on-line
learning vanish. In this case the sub-optimal fixed point is unstable but
the Jacobian is zero at the fixed point (in the 1-dimensional case)
indicating that one must go to
higher order to describe the dynamics. Standard methods for describing
the dynamics of on-line algorithms have all been developed in the
neighborhood of fixed points with negative eigenvalues and are not
applicable in this case~\citep{heskes93}. Furthermore, stability of the
fixed point is induced by fluctuations. This is contrary to our
intuition that fluctuations may be beneficial, resulting in quicker escape
from sub-optimal fixed points.  In the present case one has precisely
the opposite effect: stochasticity stabilises an otherwise unstable fixed point. In similar
studies of on-line PCA~\citep{biehl98} and back-propagation
algorithms~\citep{biehl95,saad95a,saad95b} sub-optimal fixed
points have been found which are also stabilised when the learning
rate exceeds some critical value. However, the scale of critical learning rate
stabilising these fixed points is typically $O(N^{-1})$, much
larger than in the present case. Also, the resulting time-scale for learning 
is $O(N^2)$ with a very small prefactor (in practice an $O(N)$ term will
dominate for realistic $N$). These fixed points
reflect saddle points in the mean flow while here we have a flat region
and escape is through much weaker higher order effects. This type of
sub-optimal fixed point is more reminiscent of those which have been
found in studies of small
networks, which often have a much more dramatic effect on learning 
efficiency~\citep{heskes98}. 

It is presently unclear whether on-line ICA algorithms based on Maximum-likelihood and
Information-theoretic principles~\citep[see, for
example,][]{amari96,bell95,cardoso96} exhibit
sub-optimal fixed points similar to those studied here. These
algorithms estimate a square de-mixing matrix
and will require a different theoretical treatment than for the
projection model considered here, since there may be no simple
macroscopic description of the system for large $N$.

\subsection*{Acknowledgements}

This work was supported by an EPSRC award (ref. GR/M48123).

\bibliography{ica}

\appendix 

\section{Derivation of the dynamical equations}
\label{app_dynamics} 

From equation~(\ref{eqn_dW}) one can calculate the change in $\bm R$
and $\bm Q$ (defined in~(\ref{eqn_RQ})) after a single learning step,
\begin{eqnarray}
	\Delta \bm R & = & \eta\bm\sigma \bm\phi(\bm y)\bm s^\tr + \alpha(\bm I-\bm Q)\bm R \ , \nonumber \\
	\Delta \bm Q & = & \eta\bm\sigma(\bm I+\alpha(\bm I-\bm
	Q))\bm\phi(\bm y)\bm y^\tr + \eta\bm\sigma\bm y\bm\phi(\bm
	y)^\tr(\bm I+\alpha(\bm I-\bm Q)) \nonumber
 \\ & & \ \ + 2\alpha(\bm I-\bm Q)\bm Q + \alpha^2(\bm I-\bm
	Q)^2\bm Q + \eta^2\bm\phi(\bm y)\bm x^\tr\bm x\bm\phi(\bm y)^\tr \ .
\end{eqnarray}
Here, the definition in equation~(\ref{eqn_x}) and the constraint in
equation~(\ref{eqn_Ac2}) have been used to set $\bm x^\tr \bm A_s = \bm
s^\tr$. One can obtain a set of differential equations in the limit
$N\rightarrow\infty$ using a statistical mechanics formulation which
has previously been applied
to the dynamics of on-line PCA algorithms~\citep{biehl94,biehl98} as
well as other unsupervised
and supervised learning algorithms~(see, for example,
\citet{biehl95,saad95a,saad95b} and contributions in \citet{saad98}). To obtain
differential equations one should scale the parameters of the learning
algorithm in an appropriate way, in particular $\eta\equiv \mu/N$. Typically 
one chooses $\alpha= O(1)$ but in
order to obtain an analytical solution it is more convenient to choose $\alpha\equiv
\alpha_0/N$ before taking $N\rightarrow\infty$ and then take the limit
$\alpha_0\rightarrow\infty$. The dynamics do not appear to be sensitive
to the exact value of $\alpha$ as long as $\alpha\gg\eta$ and it is
therefore hoped that the
dynamical equations are valid for $\alpha= O(1)$ which is usually
the case. The learning
rate is taken to be constant here but the dynamical equations are also valid when the
learning rate is changed slowly, as suggested for the annealed
learning studied in section~\ref{sec_asympt}. 

As $N\rightarrow\infty$ one finds,
\begin{eqnarray}
	\frac{\rd \bm R}{\rd \tau} & = & \mu\bm\sigma\langle \bm\phi(\bm y)
	\bm s^\tr \rangle + \alpha_0(\bm I-\bm Q)\bm R \ , \label{eqn_dRdtapp} \\
	\frac{\rd \bm Q}{\rd \tau} & = & \mu\bm\sigma\langle \bm\phi(\bm y)\bm
	y^\tr + \bm y\bm\phi(\bm y)^\tr\rangle + \mu^2\langle\bm\phi(\bm y)\bm\phi(\bm y)^\tr\rangle + 2\alpha_0\bm Q(\bm I-\bm
	Q),
\end{eqnarray}
where $\tau\equiv t/N$ is a rescaled time parameter. The angled
brackets denote averages over $\bm y$ as defined in
equation~(\ref{eqn_y}). In deriving the
above equations one should check that fluctuations in $\bm
R$ and $\bm Q$ vanish in the limit $N\rightarrow \infty$. This relies on 
an assumption that $\bm R= O(1)$ which may not be appropriate in some cases. For example, 
in section~\ref{sec_escape} a sub-optimal fixed point is analysed where it is 
more appropriate to consider $\bm R= O(1/\sqrt{N})$ 
and a more careful treatment of fluctuations is required.

As $\alpha_0$ is increased, $\bm Q$ approaches $\bm I$. If one sets $\bm
Q-\bm I \equiv \bm q/\alpha_0$ and make the {\it {\'a} priori} assumption that $\bm q=
O(1)$ then,
\begin{equation}
	\frac{1}{\alpha_0}\frac{\rd\bm q}{\rd \tau} = \mu\bm\sigma \langle \bm\phi(\bm y)\bm
	y^\tr + \bm y\bm\phi(\bm y)^\tr\rangle + \mu^2\langle\bm\phi(\bm
	y)\bm\phi(\bm y)^\tr\rangle -2\bm q + O(1/\alpha_0) \ .
\end{equation}
As $\alpha_0\rightarrow\infty$ one can solve for $\bm q$,
\begin{equation}
	\bm q = \frac{1}{2}\left(\mu\bm\sigma \langle \bm\phi(\bm y)\bm
	y^\tr + \bm y\bm\phi(\bm y)^\tr\rangle + \mu^2\langle\bm\phi(\bm
	y)\bm\phi(\bm y)^\tr\rangle\right) \ ,
\end{equation}
which is consistent with the {\it {\'a} priori} assumption. Substituting
this result into equation~(\ref{eqn_dRdtapp}) leads to
equation~(\ref{eqn_dRdt}) in the main text. This is an example of
adiabatic elimination of fast variables~\citep[section 6.6]{gardiner}
and greatly simplifies the dynamical equations.

\section{Eigensystem of asymptotic Jacobian}
\label{app_eigensystem}

The Jacobian of $\rd \bm R/\rd \tau$ as $\mu\rightarrow 0$ is defined
(divided by $\mu$),
\begin{equation}
    	J_{ijkl} = \left .\frac{\partial}{\partial R_{kl}}
    	\left(\left .\frac{1}{\mu}\frac{\rd R_{ij}}{\rd \tau}\right|_{\mu=0}\right)\right|_{\bm R=\bm
    	R^\ast} \ .
\end{equation}
This is a tensor rather than a matrix because the system's variables
are in a matrix. One
can think of pairs of indices $(i,j)$ and $(k,l)$ as each representing a single
index in a vectorised system. If the dynamics is equivalent
to gradient descent on some potential function
then the above quantity is proportional to the (negative) Hessian of this cost
function. The Jacobian is not guaranteed to be symmetric in the
present case, so this will not be possible in general. From
equation~(\ref{eqn_dRdt}) one obtains,
\begin{equation}
	J_{ijkl} = -\delta_{ik}\delta_{jl}\left(\xi_i +
	\mbox{$\frac{1}{2}$}\xi_j\right) -
	\mbox{$\frac{1}{2}$}\delta_{il}\delta_{jk}\xi_i \ ,
\label{eqn_J}
\end{equation}
with,
\[
	\xi_i = \left\{\begin{array}{l} \sigma_{ii}\langle s_i\phi(s_i) -
	\phi'(s_i)\rangle \ \mbox{ for } \ i \leq \min(K,M) \ , \\
        0 \ \mbox{ otherwise } \ . \end{array}\right .
\]

One must solve the following eigenvalue problem,
\begin{equation}
	\sum_{kl} J_{ijkl} V_{klnm} = \lambda_{nm} V_{ijnm} \ ,
\end{equation}
where $\lambda_{ij}$ and $V_{klij}$ are the eigenvalues and
eigenvectors respectively. A solution is required for all
$i\leq K$ and $j\leq M$ in order to get a complete set of
eigenvalues,
\begin{eqnarray}
	\lambda_{ii} & = & -2\xi_i \ , \quad V_{klii} =
	\delta_{ik}\delta_{il} \ , \nonumber \\
	\lambda_{ij} & = & -\mbox{$\frac{1}{2}$}(\xi_i+\xi_j) \ , \quad V_{klij} = 
\delta_{ik}\delta_{jl} - \delta_{jk}\delta_{il} \quad \mbox{ for
	$i<j\leq K$ } , \nonumber \\
	\lambda_{ij} & = & -\xi_i \ , \quad
	V_{klij} = \delta_{ik}\delta_{jl} \quad \mbox{ for
	$j>K$ } , \nonumber \\
	\lambda_{ij} & = & -(\xi_i+\xi_j) \ , \quad  V_{klij} = 
\xi_i \delta_{ik}\delta_{jl} + \xi_j\delta_{jk}\delta_{il} \quad
	\mbox{ for $i>j$ } .
\label{eqn_evals}
\end{eqnarray}

\end{document}